\documentclass{elsarticle}
\usepackage{amssymb}
\usepackage{amsfonts}
\newtheorem{Theorem}{Theorem}
\newtheorem{Corollary}{Corollary}
\newtheorem{Definition}{Definition}
\newtheorem{Example}{Example}

\newtheorem{Lemma}{Lemma}

\usepackage{pifont}
\def\G1{\hbox{$\displaystyle{\mbox{\ding{172}}}$}}

\begin{document}
\title{Cellular Automata Using Infinite Computations}
\author{Louis D'Alotto\\
Department of Mathematics and Computer Science\\ 
York College/City University of New York \\
Jamaica, New York 11451\\
and\\
The Doctoral Program in Computer Science\\
CUNY Graduate Center\\
{\it ldalotto@york.cuny.edu}} 
%\maketitle

\begin{abstract}
This paper presents an application of the Infinite Unit Axiom, introduced by Yaroslav Sergeyev, (see \cite{Ser1} - \cite{Ser4}) to the development of one-dimensional cellular automata. This application allows the establishment of a new and more precise metric on the space of definition for  one-dimensional cellular automata, whereby accuracy of computations is increased.  Using this new metric, open disks are defined and the number of points in each disk is computed.   The forward dynamics of a cellular automaton map are also studied via defined equivalence classes.  Using the Infinite Unit Axiom, the number of configurations that stay close to a given configuration under the shift automaton map can now be computed.   
\end{abstract}

\maketitle

\section{Introduction}

Unlike natural systems, which eventually evolve to maximal entropy, cellular automata are discrete dynamical systems that are known for their strong modeling and self-organizational properties. Defined on an infinite lattice (in the usual one-dimensional case, an infinite sequence defined on the integers), even starting with complete disorder, evolution of cellular automata can generate organized structure.   Originally developed by Von Neuman in the 1940's to model biological self-reproduction, cellular automata have long been used in computational, physical, and biological applications.  For a more complete description of applications of cellular automata see \cite{CD, AL, Springer, HM, GS, SS, Trun, W4}, and \cite{WS}.   As with all dynamical systems, it is interesting to understand the long term behavior under forward time evolution and achieve an understanding or classification of the system.  Cellular automata can be defined for any dimension greater than or equal to one.  This paper is concerned with one-dimensional or linear cellular automata defined on the integers (when no confusion arises, we will refer to one-dimensional cellular automata as simply cellular automata).   See figure 1 for an example of a one-dimensional cellular automaton.  This is Wolfram binary rule $20$ with range equal to 2, see \cite{W1} and \cite{W2}.   Starting with a random initial configuration, the cellular automaton evolves (under repeated applications) downward.  In this example, it's interesting to note the two persisting structures that emerge.  The structure on the left evolves straight down, while the structure on the right evolves down on an angle and can eventually ``crash'' into other persisting structures.  

The concept of classifying cellular automata was initiated by Stephen Wolfram, see \cite{W2}. In \cite{W2}, one-dimensional cellular automata are partitioned into four classes depending on their dynamical behavior.  Figure 1 (rule $20$) is an example of a Wolfram class $4$ cellular automaton.  A later and more rigorous classification scheme, see \cite{G1}, was developed by Robert Gilman. Here a probabilistic/measure theoretic classification was developed based on the probability of choosing a sequence that will stay arbitrarily close to a given initial sequence under forward evolution (iteration).    Gilman uses a metric that considers the central window where two sequences (configurations) agree and continue to agree upon forward iterations of a cellular automaton map.  However, in the development, this metric is limited because it doesn't take into account configurations that agree on an infinite interval to the right (or respectively to the left).  Indeed, the metric considers the absolute value of the first integral place where configurations disagree and uses that as their distance apart, see \cite{Devaney} and \cite{G1}.  For example, if configurations agree on the right hand side out to infinity but disagree in some finite position on the left, their distance apart is determined by where they disagree on the left.  In this paper, the definition of cellular automata and the metric involved are extended to include configurations that do not necessarily agree on a finite central window symmetric around $0$, and also which can agree on not necessarily symmetric infinite intervals.    

\begin{figure}
\begin{center}
\fbox{\includegraphics[width=4in,height=2.35in]{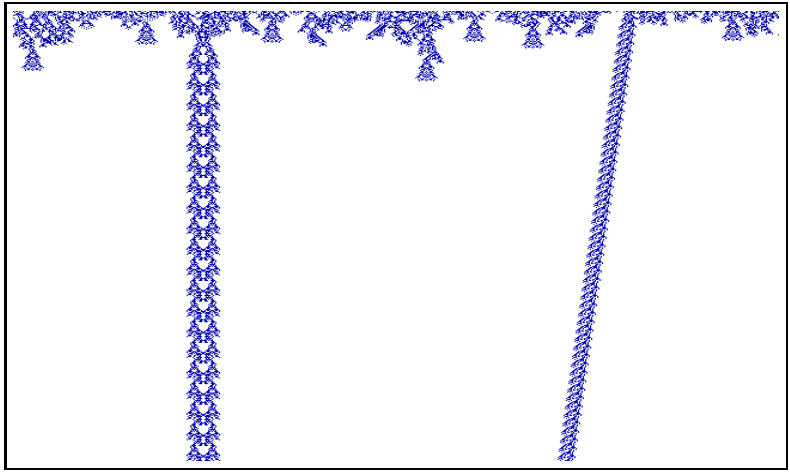}}
%\fbox{\includegraphics[width=2in,height=1.35in]{Rule20.eps}}
\caption{Example of a One-dimensional Cellular Automaton (Wolfram rule 20)}
\end{center}
\end{figure}

The classical concept of infinity has presented limitations in computations.  Indeed, metrics used on infinite sequences, and hence cellular automata, either do not allow us to observe minute differences or can lead to calculations beyond finite computations.  Analogous to the Hamming distance for finite sequences, the following metric is used to compute distances between infinite sequences.
\begin{equation} \label{equ_a} d(x,y) = \sum^\infty_{i=-\infty}\frac{|x(i)-y(i)|}{2^{|i|}}\end{equation}
Here the differences in the respective sequence values are computed and divided by $2^{|i|}$ to ensure convergence.  However, this procedure can lead to a calculation beyond finite computation and to possible inaccuracies.  For instance, using the binary alphabet S=\{0,1\}, suppose two sequences agree completely on the left of and at the $0^{th}$ place, and disagree elsewhere.  That is, they disagree on the right of $0$ or for integral values $i > 0$.  Applying the traditional well known formula to (\ref{equ_a}) yields \begin{equation}\label{equ_b} \sum^k_{i=1}\frac{1}{2^{i}}=1-\frac{1}{2^k}\end{equation}  and taking limits as $k$ approaches infinity, results in a value of $1$.  By using the {\it Infinite Unit Axiom}, see \cite{Ser1} - \cite{Ser4}, and $|\mathbb{N}|=\G1$ (the symbol $\G1$ is called grossone by Sergeyev and represents the number of elements in the set of natural numbers), the computational limitations caused by sequences that agree out to one-sided infinity or that are subject to infinite computations are overcome.  As shown for infinite $k$, that is for $k=\G1$ 
\begin{equation}\label{equ1}\sum^{\G1}_{i=1}\frac{1}{2^{i}}=1-\frac{1}{2^{\G1}}\end{equation} 
and note that $\frac{1}{2^{\G1}}$ is infinitesimal (see \cite{Ser1}).  Hence (\ref{equ1}) produces a more precise representation of the infinite sum computation.

%In a previous classification work (see \cite{G1}), a classification was built by considering the central window where two sequences agree and continue to agree upon forward iterations of a cellular automata map.  In these classifications, the central window was chosen because this is the section viewable on a computer screen.  

Before defining cellular automata with the infinite metric we need a few preliminaries.  The set of integers is denoted by $\mathbb{Z}$; $\mathbb{N}$ is the set of natural numbers and let $\mathbb{N}_0 = \mathbb{N} \cup \{0\}$.   Given a finite alphabet $S$ with two or more symbols, i.e. $|S|\geq2$, consider the space of all functions from the integers to the finite alphabet, i.e. $S^\mathbb{Z}$.  This space may also be considered as the space of all bi-infinite sequences, defined on the integers, with values taken from the alphabet $S$.  In \cite{G1} the following metric was used on the space $S^\mathbb{Z}$.  Let 
\begin{equation} \label{equ2} d(x,y)=2^{-n}, \;\; where\; n=\inf\{|i|\; : x(i) \neq y(i)\} \end{equation}  
It is noted that this metric satisfies the ultrametric property.  A metric $d(x,y)$ satisfies the ultrametric property iff it is a metric and obeys the inequality:     \[d(x,y) \leq \max \left[d(x,z),d(z,y)\right] \]   A space that satisfies the ultrametric inequality is also called a nonarchimedean space, see \cite{NBB}.  It is obvious that the triangle inequality is implied by the ultrametric inequality.   As can be seen, the metric defined in (\ref{equ2}) above considers only the symmetric window around $i = 0$.  Two sequences may agree in more integral values on one side but this information is not communicated by the metric.  In this paper this inaccuracy is overcome by using the Infinite Unit Axiom, $|\mathbb{N}| = \G1$, and developing new machinery to work with infinite sequences.  
It is noted that, in a nonarchimedean space given two open disks either one contains the other or they intersect trivially.  

%Perhaps more importantly, in a nonarchimedean space given \[\lim_{n\rightarrow \infty} a_n = 0\] implies that \[\sum^{\infty}_n a_n\] converges.   Hence there is a facilitation of the notion of convergence.   The next section gives a rigorous development of the nonarchimedean metric space used in the rest of the paper. 

% It is well known that the space $S^\mathbb{Z}$ is uncountably infinite.  However, applying the Infinite Unit Axiom, the number of sequences in $S^\mathbb{Z}$ is exactly $|S|^{2\textcircled{1}+1 }$.

\section{Metric Development}

The following definitions, lemmas and theorem will be used to develop the
framework for a generalized nonarchimedean metric on the space of
bi-infinite sequences.

\begin{Definition}
A $LTSL$ (Lower Tree Semi-Lattice) is a Lower Semi-Lattice with lower
element $\ast $ having the \textit{Semi-Total Ordering} property. A Lower
Semi Lattice has the binary operator infimum $(\wedge )$ such that the
following laws hold:

\begin{enumerate}
\item  Idempotent law $a\wedge a=a$

\item  Commutative law $a\wedge b=b\wedge a$

\item  Associative law $(a\wedge b)\wedge c=a\wedge (b\wedge c)$

\item  Lower element $*$ obeys the lower unit law $*\wedge a=*$
\end{enumerate}

The Semi-Total Ordering property is: If $x\wedge y=a$ and $x\wedge z=b$ then
either $a\leq b$ or $b\leq a$.
\end{Definition}

\begin{Lemma}
\label{lem1} In a $LTSL$, if $x\wedge y=a$, $x\wedge z=b$ and $y\wedge z=c$
then the set of elements $\{a,b,c\}$ is a totally ordered set.
\end{Lemma}

\noindent {\bf Proof:} By definition of a $LTSL$, say that $a\leq b$. Now $a\leq c$
or $c\leq a$. If the latter is true then we are done, since in this case $%
c\leq a\leq b$ holds true. Otherwise, $a\leq c$ and $c\leq b$ or $b\leq c$.
So $a\leq c\leq b$ or $a\leq b\leq c$ holds.  $\;\;\; \Box$
%\end{Proof}

\begin{Lemma}
\label{lem2} In a $LTSL$, if $x\wedge y=a$, $x\wedge z=b$, and $y\wedge z=c$
then the smallest two elements from $\{a,b,c\}$ are equal.
\end{Lemma}

\noindent {\bf Proof:}  From lemma $1$ $\{a,b,c\}$ is totally ordered. Without loss
of generality, say $a\leq b\leq c$, we will show $b\leq a$. Using $x\wedge
y\wedge z=(x\wedge z)\wedge (y\wedge z)\leq x\wedge y$ gives $b\wedge c\leq
a $, but $b\wedge c=b$, hence $a=b. \;\;\; \Box$

\begin{Definition}
An \textit{Evaluation function }is defined by\textit{\ }$F:X\rightarrow
(0,1] $, with $X$ a $LTSL$, $F(\ast )=1$, and $F$ monotonically decreasing
(i.e. $a\leq b$ implies $F(a)\geq F(b)$).
\end{Definition}

\begin{Theorem}
If $d:X\times X\rightarrow \lbrack 0,1]$, with $X$ a $LTSL$, and $F$ is an
evaluation function where $d(x,y)=F(x\wedge y)$ whenever $x\neq y$, and $%
d(x,y)=0$ otherwise, then $d$ is a nonarchimedean metric.
\end{Theorem}
\setcounter{Proof}{0}
{\bf Proof:} By construction $d$ is positive definite, and symmetric. The
ultrametric triangle inequality: $d(x,y)\leq \max [d(x,z),\,d(z,y)]$ holds
since if two or more elements in $\{x,y,z\}$ are equal then at least one
term from $d(x,y)\leq \max [d(x,z),\,d(z,y)]$ is zero and the others are
equal to each other. Otherwise $F(x\wedge y)\leq \max \left[ F(x\wedge
z),F(z\wedge y)\right] $ holds true and applying Lemma \ref{lem2} and the
non-increasing property of $F$, it follows that the two largest terms among $
F(x\wedge y)$, $F(x\wedge z)$, and $F(z\wedge y)$ must be equal.$\quad \Box $
%\end{Proof}

A nonarchimedean metric $d$ is thus defined on a $LTSL$ $X$, for $f,g\in X,$
as follows: 
\[
d(f,g)=\left\{ 
\begin{array}{ll}
0 & if\quad f=g \\ 
F(f\wedge g) & otherwise
\end{array}
\right. 
\]

Let $S$\textbf{\ }be a finite alphabet of size $s \geq 2$ and let $X$ = $(S\cup \{*\})^{\mathbf{Z}}$. $X$ is the set of all maps from
the integers to $S\cup \{*\}$. That is, for $x\in X$, $x:\mathbb{Z}\rightarrow S\cup \{*\}$. It is noted that the set $S\cup \{*\}$ is compact and hence the product space $X $ is also compact.  As the next definitions show, the Infinite Unit Axiom will be applied and used in the construction of the metric for computations with infinite sequences.  The above lemmas and theorem show that the constructed metric is an ultrametric and the space is nonarchimedean. 

{\Definition Let \[x \wedge y = \left\{ 
\begin{array}{ll}
x & if \;\; x=y \\ 
{*} & if\;\; x(0) \neq y(0) \; or \; x(0) = * \\
x(m) ...x(0)...x(n) & if \;\;x(i) = y(i) \forall i \in [m,n]  \;\; and \; * outside 
\end{array}
\right. 
 \]
}
Note:  $m\leq 0$ and can equal $-\G1$, similarly $n\geq 0$ can equal $\G1$. Hence computations on infinite sequences are allowed.  Thus, $x \wedge y$ is the place where two sequences agree on the largest stretch around $0$ and is $*$ valued outside.    

{\Definition \[F(x \wedge y) = \left\{ 
\begin{array}{ll}
1 & if \;\; x \wedge y = {*}\\ 
2^{-(n+1-m)}& if\;\; x \wedge y = ...***x(m)...x(0)...x(n)***...
\end{array}
\right. 
 \]
 }
We form the following metric on the space of bi-infinite sequences: 
\[d(x,y) = \left\{ 
\begin{array}{ll} 
0 & if \;\; x = y \\
F(x\wedge y) & otherwise 
\end{array}
\right. \]

{\Example Given $S = \{0,1\}$, let $x = ...111\langle1\rangle111...$ and $y= ...00011\langle 1 \rangle 111...$ 
In our examples, when not explicitly denoted, we will use the symbol $\langle \; \rangle$ to denote the zeroth place.  The sequences $x$ and $y$ agree completely on the right hand side, and at integral values $0, -1$, and $-2$.  
 \[x\wedge y = ...***x(-2)x(-1)x(0)....x(n)....x(\G1)\] } \[F(x \wedge y) = 2^{-(\G1+1-(-2))}\] and \[d(x,y)=\frac{1}{2^{\G1+3}}\] 
Hence, the distance between the two points $x$ and $y$  is infinitesimal.  As the following example shows, the above construction  easily covers the finite distance case. 
%%%%%%%%%%%%%%%%%%%%%%%%%%%%%%%%%%%%%%%%%%%%%%%%%%%

{\Example Again, using the binary alphabet $S=\{0,1\}$, let $x = ...1110\langle1\rangle0111...$ and $y = ...1100\langle1\rangle0101...$ \[x\wedge y = ...***x(-1)x(0)x(1)x(2)***...\] That is, the sequences differ in the $-2$ and $3^{rd}$ integral positions. Hence,  \[F(x \wedge y) = 2^{-(2+1-(-1))} = 2^{-4}\] and \[d(x,y) = \frac{1}{2^4}\]}

%%%%%%%%%%%%%%%%%%%%%%%%%%%%%%%%%%%%%%%%%%%%%%%%%%% 

 \section{Cellular Automata}
 %%%%%%%%%%%%
As before, $S$ is an alphabet of size $s$ such that  $s \geq 2$ and let $X=S^{\mathbb{Z}}$, i.e. the set of all maps from the lattice $\mathbb{Z}$ to the set  $S$.  That is, for $ x \in X$, $x: \mathbb{Z} \rightarrow  S$.  Cellular automata are induced by arbitrary (local) maps: \[F: S^{(2r+1)} \longrightarrow  S \]
These are usually called local rules or block maps in the literature, see \cite{GAH} and \cite{G1}.  The value $r \in \mathbb{N}_0$ is called the range of the map.  The automaton map $f$ induced by $F$ is defined by $f(x)=y$ with \[y(i)=F[x(i-r),...,x(i+r)]\]
To illustrate the importance of discrete time steps in the forward evolution of the automaton, we will use the following formula where $t$ represents time.
\[y(i)_{t+1}=F[x(i-r)_t,...,x(i+r)_t]\]
The restriction of $x \in X$ to a non-empty interval $[i,j]$ of $\mathbb{Z}$, where $-\G1 \leq i \leq j \leq \G1$ is called a {\it word}.  Words are written $x[i,j]$. The length of a word $w = x[i,j]$ is $|w| = j-i+1$.  It is important to note that, using $\G1$, words (or the length of a word) can be infinite, however cannot have an endpoint greater than $\G1$ (nor less than $-\G1$).   Also, for any $a \in S$, define $x_a \in X$ by $x_a(i)=a$, for $i \in \mathbb{Z}$. 
 %%%%%%%%%%%%
 Under the usual product topology, a {\it cylinder} is a set $C(i,j,w) = \{x \in X | x[i,j] = w\} $, where $|w|=j-i+1$.  Note that cylinders are both open and closed. We define the open disk of radius $2^{-(n-m)}$ around $x$ to be $C_{[m,n]}(x) = C(m,n,x[m,n])$ for $m \leq 0 \leq n$.  It will be shown that the number of points in an open disk can now be determined.  

The following is a simple, but important, example of a cellular automaton of range $r = 1$.  The evolutionary behavior of this automaton is clearly exhibited.    
\begin{Example}
Let $ S = \{0,1\}$ and let $f$ be the automaton induced by the local rule $F: S^3 \rightarrow  S$ by $F(1,1,1)=1$ and $F(a,b,c)=0$ otherwise.  If we apply forward iterations of the induced automaton map $f$, all sequences eventually go to the quiescent state of $x_0$, except for the initial sequence $x_1$ which remains constant.
\end{Example}  
In the previous example, given any finite word $x[i,j]$ with at least one element in the word not equal to $1$, and an open disk $C_{[i,j]}(x)$ around it, every point in the open disk $C_{[i,j]}(x)$, will eventually evolve, under forward iterations, to the quiescent state of $x_0$.  Hence it makes sense to determine how many elements are in these open disks.  Theorem $2$ and its corollary below will answer this question.  There are numerous other examples of cellular automata maps.  A more chaotic rule can be seen via the following example.
\begin{Example}
Let $ S = \{0,1\}$ and let $f$ be the automaton induced by the local rule $F: S^3 \rightarrow  S$ by $F(a,b,c)= (a+c)mod \; 2$.  Applying forward iterations of the induced automaton map $f$ yields no particular pattern.  Beginning with an initial random configuration in $S = \{0,1\}^\mathbb{Z}$ can yield many different configuration sequences.  
\end{Example}

 %To understand the dynamics of cellular automata it is necessary to study the classes of forward iterates of configurations (sequences) that stay ``close" to those of a given sequence.  Here we define an equivalence relation $x \approx y$ iff $\forall i \in \mathbb{N}_0$, $d(f^{i}(x),f^{i}(y)) < 2^{-|n+1-m|}$.  The equivalence classes are $B_{m,n}$.  That is,  \[B_{m,n}(x)=\{y\;|\; d(f^i(y),f^i(x)) < 2^{-|n+1-m|} \;\; \forall i \in \mathbb{N}_0\}\] 
% $B_{m,n}(x)$ is the set of $y$ for which $(f^{i}(y))[m,n] = (f^{i}(x))[m,n]$, $\forall i \in \mathbb{N}_0$.
 
 %{\Definition A non-empty word $y[m,n]$ is ultimately periodic if there exists $k \geq 0$, $p \leq 0$ and $q \geq 0$ such that $(f^{n+i}(y))[p,q] = (f^{i}(y))[p,q]$ for some $n \geq 1$ and all $i \geq k$.  The smallest $n$ such that $(f^{n+i}(y))[p,q] = (f^{i}(y))[p,q]$ is called the period of the word.}
 
% The forward orbit of $x$ under the automaton map $f$ will be denoted by $O^+(x)$.  Similarly, the backward orbit of $x$ will be denoted by $O^-(x)$.  $\overline{O^+(x)}$ is the closure of $O^+(x)$ and for any $A \subset X$, $A^\G1$ is the interior of $A$.
 
 %%%%%%%%%%%%%%%%%% Number of Elements in S_Z %%%%%%%%%%%%%%%%%
 \begin{Theorem} Given the space $S^\mathbb{Z}$ of bi-infinite sequences, the number of elements $x \in S^\mathbb{Z}$ is equal to $|S|^{2\G1+1}.$
 \end{Theorem}
   
\noindent {\bf Proof:} $|S|$ is the number of elements of the finite alphabet $S$.  Over the integers, $\mathbb{Z}$, there are exactly $\G1$ elements to the right and exactly $\G1$ elements to the left of $0$.  Hence there are $|S|^{\G1}$ choices in the sequence to the right and  $|S|^{\G1}$ choices in the sequence to the left of $0$, plus $|S|$ choices at the $0th$ place.  Therefore $|S^\mathbb{Z}| = |S|^{\G1} \cdot |S|^{\G1} \cdot |S| = |S|^{2\G1+1}. \;\;\; \Box$

The proof of the following corollary is similar to that of the previous theorem and hence omitted.
{\Corollary The open disk $C_{[m.n]}(x)$ around $x$ contains $|S|^{2\G1-(n-m)}$ elements.}

\vspace{4mm}

As mentioned in the introduction of this paper, Wolfram, \cite{W1}, \cite{W2}, observed, after extensive computer simulations, that the dynamics of a cellular automaton map can behave differently under different initial configurations.  However, it was also observed that if the configuration was chosen at random, the probability is high that the dynamical behavior of the automaton map $f$ will lie in one of four classes.  In \cite{G1} Gilman shows that a cellular automaton must lie in one of three classes.  These classes are determined by the probability of finding another configuration that stays arbitrarily close to an initial configuration.  Hence it makes sense to study cellular automata by determining the number of configurations that stay within a certain distance of the initial configuration under the automaton map.  

%accompanied by a fixed probability measure, $\mu$.   

To understand the dynamics of cellular automata it is necessary to study the forward iterates of configurations that stay ``close" to those of a given configuration, call it ``$x$''.  $f^i(x)$ is used to represent the $i^{th}$ iterate of the automaton function $f$.  That is, $f^i(x)=f\circ f \circ f \cdot\cdot\cdot \circ f (x)$.   Here an equivalence relation $x \approx y$ iff $\forall i \in \mathbb{N}_0$, $d(f^{i}(x),f^{i}(y)) < 2^{-|n+1-m|}$ is defined.  The equivalence classes are $B_{m,n}(x)$.  That is,  \[B_{m,n}(x)=\{y\;|\; d(f^i(y),f^i(x)) < 2^{-|n+1-m|} \;\; \forall i \in \mathbb{N}_0\}\] 
$B_{m,n}(x)$ is the set of $y$ for which $(f^{i}(y))[m,n] = (f^{i}(x))[m,n]$, $\forall i \in \mathbb{N}_0$.  Recall, $(f^{i}(y))[m,n]$ represents words and that $m \leq 0 \leq n$.

\begin{Example}
The left shift map, $\sigma$, is a cellular automaton of range $1$, defined by $\sigma(x_{i}) = x_{i+1}$.  i.e. the map that shifts all symbols of a configuration to the left, as illustrated below:  

\[\begin{array}{lll} \;\;\;\;x & = & ...0\;1\;1\;1\;0\;0\;1\;1\;0\;1\;1\; \langle1\rangle\;0\;1\;0\;0\;1\;0\;1\;0\;0\;0\;1\;1... \\
                                \;  \sigma(x) & = & ...1\;1\;1\;0\;0\;1\;1\;0\;1\;1\; 1\; \langle0\rangle\;1\;0\;0\;1\;0\;1\;0\;0\;0\;1\;1 \;...  \\
                                  \sigma^2(x) & = & ...1\;1\;0\;0\;1\;1\;0\;1\;1\;1\; 0\; \langle1\rangle\;0\;0\;1\;0\;1\;0\;0\;0\;1\;1 \; \;...  \\
                                 \;\; \cdot \\
                                  \;\; \cdot \\
                                  \;\;  \cdot 
\end{array} \]
 All configurations $y \in B_{m,n}(x)$ would have to agree with $x$ to the right, out to $\G1$ and at the zeroth place.  Hence it is obvious that $B_{m,n}(x)$ contains elements.  The number of elements in $B_{m,n}(x)$ is at most $|S|^{\G1}+1$  
 
\end{Example}
The previous example can be extended for the right shift as well.  Hence it is shown that, given the right or left shift automaton map, the number of elements in $B_{m,n}(x)$ is at most $|S|^{\G1}+1$.  
 
 %%%%%%%%%%%%%%%%%%%%%%%%%%%%%%%%%%%%%%%%%%%%%%%%%%
% \section{Cellular Automata with Infinite Range}
% By definition cellular automata are induced by arbitrary maps: $F:S^{(2r+1)} \longrightarrow S$,  where the value $r \in \mathbb{N}_0$ is called the range of the map.  In the traditional definition $r$ is a non-negative integer and hence finite.  However, by assuming the Infinite Unit Axiom, $r$ can be infinite.  That is, $r$ can equal $\G1$.   To define this, the {\it extended integers} $\widehat{\mathbb{Z}}$ are needed.  

%\[\widehat{\mathbb{Z}} = \{...,-2\G1,...,-\G1-1,-\G1,-\G1+1,...,-1,0,1,...\G1-1,\G1,\G1+1,...,2\G1,... \}\] 
 
% Here the local maps are: \[F:  S^{2\G1+1} \longrightarrow S\] with range $r = \G1$. 
 %The automaton map $f$ induced by $F$, defined by $f(x)=y$ is:  \[y(i)=F[x(i-\G1), x(i-\G1 +1),...,x(i+\G1-1),x(i+\G1) ]\]
 % Note that the index $i$ is an integer and $-\G1\leq i \leq \G1$.   
 %The following, an extension of the previous example,  is a cellular automaton with infinite range: 
 %{\Example Given the local map
% \[F: S^{2\G1+1} \longrightarrow S  \]
%Defined by \[F(...,a_{-1},a_0,a_1,...) =  \left\{ \begin{array}{ll} 1 & if \; a_k=1 \;\forall k \in {\mathbb{Z}} \\											    0 & otherwise   \end{array} \right.\]
 %}
%Again, if we apply forward iterations of the induced automaton map $f$ all points except $x_1$ tend to the quiescent state.
\section{Conclusion}
In this paper, the framework for defining and working with cellular automata has been extended by applying the Infinite Unit Axiom, $|\mathbb{N}|=\G1$.  In the classical sense, the space $S^\mathbb{Z}$ is considered uncountable and beyond our computational abilities.  Moreover, in the classical sense, the open disks $C_{[i,j]}(x)$ around  a point $x$ contain uncountably many points.  By assuming the Infinite Unit Axiom, the number of points in each open disk can now be known.    
Usual metrics on the space $S^\mathbb{Z}$ (the space of definition for cellular automata) are limited in accuracy.  Indeed, sequences can agree on infinite intervals and not have this information communicated by the metric.  This inaccuracy has also been overcome by applying the Infinite Unit Axiom and developing a new metric.   In studying the dynamics of an automaton map, it is necessary to analyze the elements of the forward iterations and the quantity of elements that stay infinitesimally close to a given configuration under the map.  This was accomplished by looking at the sets $B_{m,n}(x)$ for the shift map and, using $\G1$, determining the number of elements contained within.

\end{document}